\documentclass{article}
\usepackage{spconf,amsmath,graphicx,hyperref,tabularx, booktabs, svg, pdfpages, textcomp}
\usepackage{bibspacing}

\title{Layer-Aware Early Fusion of Acoustic and Linguistic Embeddings for Cognitive Status Classification}
\name{Krystof Novotny$^{\star \dagger}$ \qquad Laureano Moro-Velázquez$^{\ddagger}$ \qquad Jiri Mekyska$^{\star \dagger}$}
  
  \address{$^{\star}$ Department of Telecommunications, Brno University of Technology, Brno, Czech Republic \\
      $^{\dagger}$Central European Institute of Technology (CEITEC), Masaryk University, Brno, Czech Republic \\
      $^{\ddagger}$ Department of Electrical and Computer Engineering, Johns Hopkins University, Baltimore, MD, USA}

\begin{document}
%
\maketitle
{

\ninept
\fontsize{9.5pt}{12pt}\selectfont

\begin{abstract}
Speech contains both acoustic and linguistic patterns that reflect cognitive decline, and therefore models describing only one domain cannot fully capture such complexity. This study investigates how early fusion (EF) of speech and its corresponding transcription text embeddings, with attention to encoder layer depth, can improve cognitive status classification. Using a DementiaBank-derived collection of recordings (1,629 speakers; cognitively normal controls--CN, Mild Cognitive Impairment--MCI, and Alzheimer’s Disease and Related Dementias--ADRD), we extracted frame-aligned embeddings from different internal layers of wav2vec 2.0 or Whisper combined with DistilBERT or RoBERTa. Unimodal, EF and late fusion (LF) models were trained with a transformer classifier, optimized, and then evaluated across 10 seeds. Performance consistently peaked in mid encoder layers ($\sim$8--10), with the single best F1 at Whisper + RoBERTa layer 9 and the best log loss at Whisper + DistilBERT layer 10. Acoustic-only models consistently outperformed text-only variants. EF boosts discrimination for genuinely acoustic embeddings, whereas LF improves probability calibration. Layer choice critically shapes clinical multimodal synergy.

\end{abstract}
\section{Introduction}
\label{sec:intro}










Pre-trained models have become a central element of speech processing, offering rich acoustic and linguistic embeddings that are increasingly used in clinical tasks. Recent studies have shown that representations of both domains can successfully support the detection of cognitive impairment. However, while these models achieve strong classification performance, their interpretability and explainability remain limited. Understanding which aspects drive decisions, and how complementary information from different modalities contributes to cognitive state classification, is an open challenge that this study seeks to address. 

Cognitive decline manifests in subtle speech cues, ranging from prosodic variation to lexical simplification and disfluencies. These cues span both the acoustic and linguistic domains, and no single modality fully captures the complexity of dementia-related speech. This motivates the hypothesis that models combining heterogeneous representations from both domains can provide better diagnostic performance than unimodal models. 

Early fusion (EF), which integrates acoustic and linguistic embeddings at the feature level prior to classification, provides an attractive framework. Unlike late fusion (LF), which aggregates model outputs, EF enables richer joint representation learning and leverages temporal or semantic complementarities. In this work, we examine embeddings from multiple pre-trained models (both acoustic and linguistic), as well as EF and LF approaches, and evaluate their effectiveness for classifying participants as cognitively normal subjects (CN), individuals with mild cognitive impairment (MCI), or those with Alzheimer's disease and related dementias (ADRD). Beyond accuracy, our study aims to probe the mechanisms by which different embeddings contribute to classification.

\subsection{State of the Art}

Across recent studies on automated speech-based dementia detection, three themes emerge. First, research moves from single-modality baselines to multimodal designs. Combining acoustic features with transcripts and ensembling models yields complementary gains, with simple score fusion being a common initial tactic \cite{campbell2021,syed2021,martinc2021}.

Second, how information is combined matters. Side-by-side evaluations report mixed outcomes for early or late combination, whereas methods that let one stream guide the other (such as cross-attention between speech and text) tend to be stronger and easier to inspect \cite{krstev2022,pan2024,altinok2024,zhang2025}.

Third, representation depth is now a key lever. Selecting or weighting layers from pre-trained speech encoders delivers competitive acoustic models and clarifies which parts of the signal drive decisions \cite{pan2025,javanmardi2024,li2024}.

Overall, the field is shifting from hand-crafted indicators to learned representations, from naive concatenation to guided interaction, and toward visualizable rationales -- hence, precisely the pivots addressed by our research.

\subsection{Research Questions}
Building on this literature, our work is motivated by the following questions:

\textbf{Multimodal synergy:} Does improved classification performance stem specifically from combining heterogeneous domains (acoustic and linguistic), or can the gains be explained by generic ensembling effects?

\textbf{Layer selection:} For acoustic models such as wav2vec\,2.0 or Whisper, is the final-layer representation optimal for cognitive assessment, or do intermediate layers provide more relevant paralinguistic features?

\textbf{Temporal grounding of linguistic features:} Since text embeddings lack timing and pause information, can integrating temporal structure (e.g., through EF with acoustic models) enhance detection performance?

\textbf{Semantic leakage in acoustic models:} To what extent do self-supervised acoustic embeddings capture linguistic information? Does enriching them with text embeddings provide complementary signals or redundancy?

By addressing these questions, we aim not only to optimize classification accuracy but also to advance understanding of how multimodal speech and language representations contribute to cognitive health assessment.

\section{Methodology}
\label{sec:format}

\subsection{Materials}
\label{sssec:subsubhead}







The experiments presented in this study were inspired by the Pioneering Research for Early Prediction of Alzheimer’s and Related Dementias EUREKA Challenge\footnote{\href{https://www.drivendata.org/competitions/299/competition-nih-alzheimers-acoustic-2/page/929/}{PREPARE Challenge official website}} by the National
Institute on Aging, which focused on developing models for automated classification of cognitive status from speech. 

All experiments were conducted exclusively on the English-speaking subset of the challenge pre-processed dataset. This dataset comprises $\sim$30-second audio clips of speech from individuals collected across multiple studies and open-access corpora~\cite{DB}. Each recording corresponds to a unique individual. Demographic metadata (age, sex and original corpora) were also available. Our experimental subset included a total of 1,629 participants. Table~\ref{tab:demo} summarizes the age and sex distributions across the analyzed cohorts.

To train and evaluate the classification models, the dataset was divided into training (64\,\%), validation (16\,\%), and test (20\,\%) sets using stratified sampling based on diagnosis, sex and original corpora, ensuring proportional representation of cognitive classes across all subsets. The partitioning was performed with a fixed seed to ensure reproducibility.


\begin{table}[ht]
\centering
\caption{Summary statistics of the cognitive state classes distribution.}
\vspace{0.5em}

\label{tab:demo}
\begin{tabularx}{\linewidth}{>
{\arraybackslash}X|>
{\centering\arraybackslash}m{7.2em}>
{\centering\arraybackslash}m{7.2em}}
\toprule
\textbf{Cognitive state} & \textbf{Count (M / F)} & \textbf{Age (mean ± std)} \\
\midrule
CN & 929 (388 / 541) & 74.9 ± 8.4 \\ 
MCI & 134 (66 / 68) & 72.5 ± 7.3 \\ 
ADRD & 566 (239 / 327) & 75.9 ± 8.3 \\ 
\bottomrule
\end{tabularx}
\vspace{-0.5em}
\vspace{1em}

\end{table}

\subsection{Audio and Text Representations}
\label{sssec:subsubhead}





To model speech using both acoustic and linguistic information, we implemented a multimodal feature extraction pipeline that enables frame-level alignment for EF of audio and text representations (see Fig.~\ref{fig:diagram})\footnote{\textcolor{black}{\href{https://github.com/BDALab/ICASSP_2026/}{The code repository for reproducing our experiments is available at: https://github.com/BDALab/ICASSP\_2026.}}}. This process was applied across four model combinations: DistilBERT~\cite{sanh2019}\footnote{\href{https://huggingface.co/distilbert/distilbert-base-uncased}{HuggingFace: distilbert/distilbert-base-uncased}} with wav2vec\,2.0~\cite{baevski2020}\footnote{\href{https://huggingface.co/facebook/wav2vec2-base}{HuggingFace: facebook/wav2vec2-base}}, DistilBERT with Whisper~\cite{radford2023}\footnote{\href{https://huggingface.co/openai/whisper-small}{HuggingFace: openai/whisper-small}}, RoBERTa~\cite{liu2019}\footnote{\href{https://huggingface.co/FacebookAI/roberta-base}{HuggingFace: FacebookAI/roberta-base}} with wav2vec\,2.0, and RoBERTa with Whisper. For each combination, audio features were extracted from all encoder layers individually, resulting in 12 distinct audio representations per combination (both selected acoustic models have 12 internal layers). Each model extracts a 768-dimensional dense vector space ($D_\text{audio}$,  $D_\text{text}$).

\textbf{Audio Representations.} Raw speech was resampled to 16\,kHz and, using wav2vec\,2.0 or Whisper, we extracted encoder-layer embeddings. Hidden states served as frame-wise features. Temporal resolution was derived from frame count and duration ($\sim$20\,ms step) and used consistently for EF alignment.


\textbf{Text Transcription and Word Alignment.} WhisperX~\cite{bain2022}\footnote{\href{https://github.com/m-bain/whisperX}{GitHub: WhisperX}} transcribed audio and produced word-level timestamps. Timestamps were converted to frame indices using the audio model’s temporal resolution. Minor postprocessing fixed boundary overlaps, ensuring non-overlapping, monotonically increasing frame spans.


\textbf{Text Representations and Tokenization.} Transcripts were tokenized and encoded with DistilBERT or RoBERTa. Embeddings were taken as last hidden states. Subword tokens received per-token start/end frames, allocated proportionally to character lengths for accurate temporal alignment.


\textbf{Token-Level Frame Mapping.} Frame-wise acoustic features were concatenated with token embeddings repeated over each token's frame span, yielding $[T, D_\text{audio} + D_\text{text}]$, where $T$ is the number of frames. The fused tensor captured both low-level acoustic properties and contextual linguistic information for frame-level learning.
\vspace{1em}


This process was applied to each recording in the dataset, yielding a collection of multimodal tensors for downstream classification tasks. All precomputed outputs were cached for each model-layer combination (i.e., the acoustic and linguistic representations themselves before fusion). 




\begin{figure*}[t]
    \centering
        \includegraphics[width=0.98\linewidth]{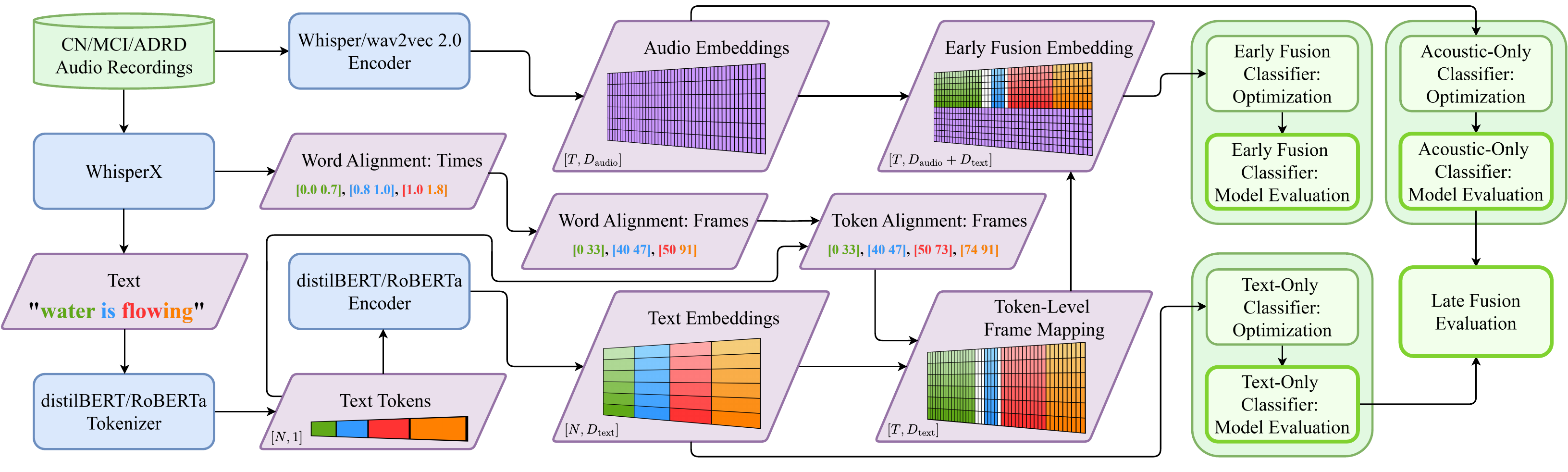}
   \vspace{-0.3em}
   \vspace{-0.3em}
\caption{Block diagram of a pipeline for EF and training of classification models. $N$ -- number of tokens, $T$ -- number of frames, $D_\text{text}$ -- size of text vector representation, $D_\text{audio}$ -- size of audio vector representation.}
  \label{fig:diagram}
\vspace{-0.4em}
\end{figure*}

\textbf{Time-Aware Modifications of the Linguistic Encoder.} We evaluated two DistilBERT/RoBERTa variants encoding word timing and pauses. TA-\,variant replaces standard positional indices with WhisperX-derived token start times mapped to 20\,ms frames (i.e., the "Token Alignment: Frames" block in Fig.~\ref{fig:diagram}). TA-PAD-\,variant inserts \textit{[PAD]} tokens between words and assigns each of them a positional index corresponding to the beginning of the inter-word silence on the same scale of 20\,ms frames.

\subsection{Classification}
\label{sssec:subsubhead}






Across scenarios, we classified extracted representations (audio, text, or EF) by assigning each recording to CN, MCI, or ADRD group. The pipeline had two phases: hyperparameter tuning, then 10 times repeated evaluation.



\textbf{Classifier architecture.} The transformer-based classifier consisted of a number of configurable stacked layers (heads, hidden size, dropout) with positional encodings. Sequence outputs were aggregated by mask-weighted mean or learnable attention pooling. The pooled vector fed a feed-forward head with dropout and softmax for 3-class prediction. The training objective followed the multiclass log loss.



\textbf{Hyperparameter optimization.} Models were tuned in Optuna using a TPE sampler. The search covered learning rate, model width/depth (layers, heads, hidden size), dropout, weight decay, batch size, projection use, pooling, positional encoding, normalization, and LR scheduling. Each trial used early stopping on validation log loss (150 trials). The best analyzed configuration was used for evaluation.

 \textbf{Evaluation.} After determining the optimal hyperparameters, the entire process (data split, model initialization, and training) was executed independently ten times for evaluation purposes, with each run using a different seed. Seeds were applied consistently across pipeline backend modules to ensure reproducibility of each run. The evaluation metrics were log loss and F1 score, computed separately for training, validation, and test partitions. We report multiclass log loss alongside F1 because it evaluates probability calibration and penalizes overconfident errors, which is crucial for clinically oriented risk estimation and threshold selection under class imbalance. For each experimental condition, the final reported performance corresponds to the mean of the ten runs.

This two-stage approach, first optimizing hyperparameters and then assessing across multiple seeds, ensured reliable estimation of generalization performance.


\textbf{Late fusion.} For each recording, we averaged class probabilities from separately trained audio- and text-only models as a LF strategy. 

\section{Results}
\label{sec:pagestyle}

The best results for EF, LF, and also for acoustic- and text-only models are presented in Table~\ref{tab:f1_results}. Figures \ref{fig:ll} and \ref{fig:f1} show the log loss and F1 score results for all experiments within one selected scenario (Whisper + RoBERTa). EF (Whisper + RoBERTa, layer~9) gave the top F1 score (0.633) overall, while the lowest log loss (0.678) was from LF (Whisper + DistilBERT, layer 10). TA- and TA-PAD- variants achieved similar or worse results than basic linguistic representations.

Across 48 settings, EF yielded the highest F1 in 81.2\,\% of cases, while LF achieved the lowest log loss in 70.8\,\%. Acoustic-only models beat text-only in all cases and nearly matched EF and LF (notably Whisper). Peak performance clustered around layers 8–10. Metrics sometimes diverged (e.g., at layer 9, LF achieved the best log loss, whereas EF had the best F1 score).

\begin{table}[ht]
\centering
\caption{Best F1 score and log loss results for each strategy with its scenario and layer noted.}
\label{tab:f1_results}
\begin{tabularx}{\linewidth}{>
{\arraybackslash}X|>
{\centering\arraybackslash}m{2.5em}>
{\centering\arraybackslash}m{2.5em}>
{\centering\arraybackslash}m{2.5em}}
\toprule
\textbf{Modeling strategy} & \textbf{Layer} & \textbf{F1 score}  & \textbf{Log loss} \\
\midrule
EF (Whisper + RoBERTa) & 9 & \textbf{0.633} & \textbf{0.687} \\
\midrule
LF (Whisper + DistilBERT)    & 9 & \textbf{0.596} & 0.679 \\
LF (Whisper + DistilBERT)     & 10 & 0.585 & \textbf{0.678} \\
\midrule
Acoustic-only (Whisper)            & 10 & \textbf{0.622} & \textbf{0.686} \\
\midrule
Text-only (TA-DistilBERT)           & — & \textbf{0.492} & 0.814 \\
Text-only (DistilBERT)         & — & 0.491 & \textbf{0.803} \\
\bottomrule
\end{tabularx}
\vspace{-1.5em}
\end{table}



\section{Discussion}
\label{sec:typestyle}

EF yields the largest gains over unimodal baselines when the acoustic encoder operates in a genuinely acoustic regime. At lower to mid layers, acoustic embeddings are less dominated by linguistic content and still carry substantial prosodic and paralinguistic cues; fusing them with linguistic embeddings boosts discrimination without redundancy. \textcolor{black}{This interpretation is consistent with layer-wise analyses of self-supervised speech encoders, which
report a progression from local acoustic cues to phonetic identity and then increasing word/semantic content as depth grows \cite{pasad2021, pasad2023}.} At higher layers, where the information stream increasingly carries lexical/semantic content, EF no longer surpasses the acoustic-only model and can even degrade performance, corresponding to an increased overlap of content-like information.

\begin{figure}[t!]

 \begin{center}
   \includegraphics[width=0.91\linewidth]{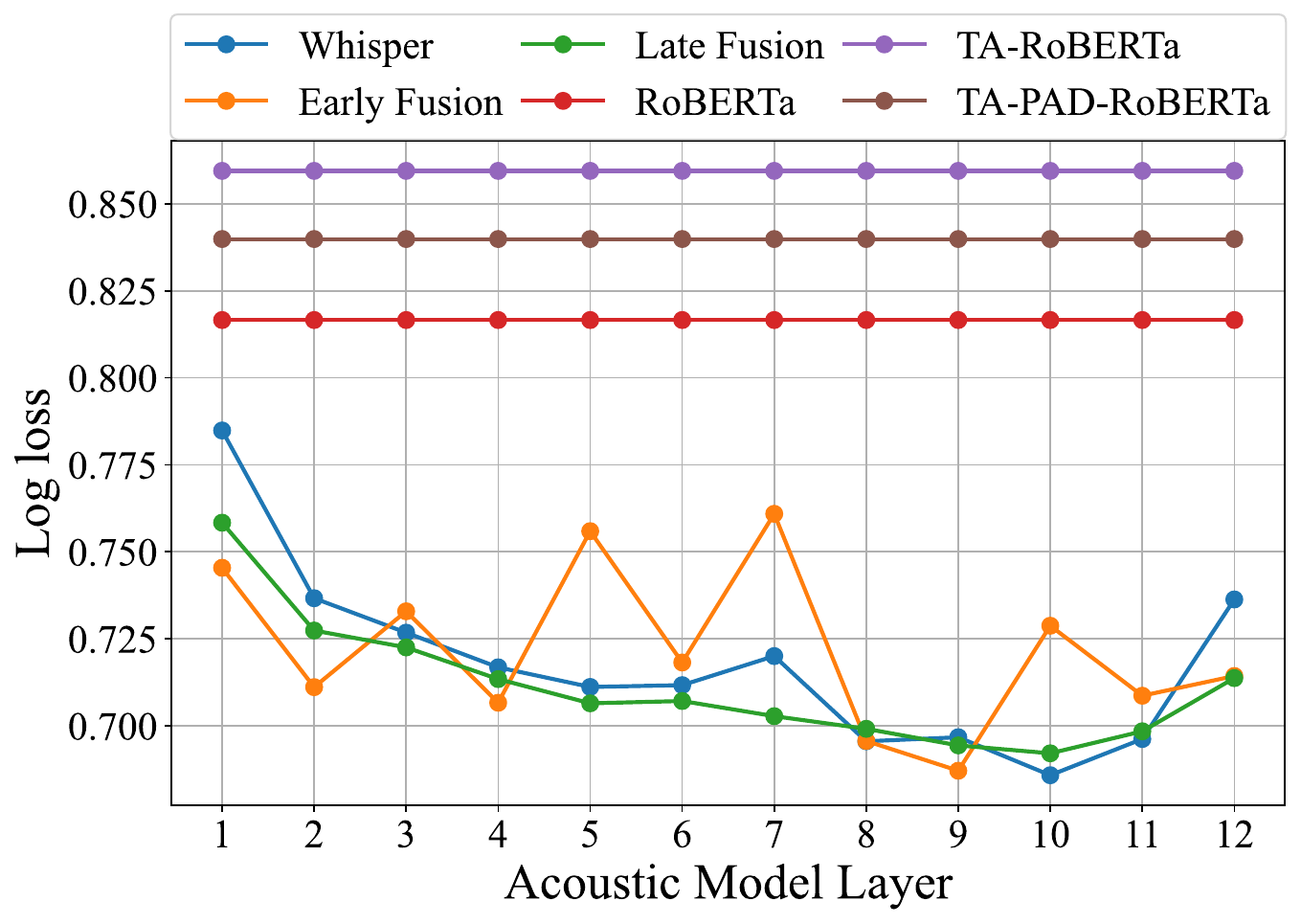} 
   \vspace{-0.1em}
   \caption{Log loss results of selected strategies depending on the acoustic model layer for Whisper + RoBERTa scenario.}
    \label{fig:ll}
\vspace{-1em}
\vspace{-1.35em}
 \end{center}

\end{figure}

Layer-wise trends were consistent: acoustic-only and thus also LF performance peaked around layers 8–10, whereas EF peaked earlier (layers 7–9). This leftward shift accords with EF’s dependence on heterogeneity. As depth increases, the acoustic representation becomes more linguistic and less complementary to text. \textcolor{black}{Such intermediate-layer optima and the frequent suboptimality of the very last layer have also been observed across different acoustic models and tasks, where single well-chosen layers can match or outperform using all layers in frozen settings \cite{pasad2023, maji2024, cooper2025}.} 

The same windows held for both F1 and log loss. We attempted to confirm this hypothesis by analyzing cosine similarities within the acoustic embeddings themselves. We did this in two cases — for the first and last layer of the model. Recurring patterns in the last layer confirmed semantic leakage. In contrast, we observed an absence of these lexical patterns in the first, therefore signal-driven, layer. \textcolor{black}{This aligns with reports that word-level (and partially semantic) structure emerges more strongly in higher layers, while lower layers remain predominantly signal-driven \cite{pasad2021, pasad2023}.} Thus, EF is best for fusing truly acoustic embeddings. Under semantic convergence, LF acts as robust probabilistic ensembling. Model-specific patterns align with this view. For Whisper, the final-layer acoustic-only model exceeded EF on F1, implying strong content integration and making other text features redundant. \textcolor{black}{This aligns with evidence that Whisper encoder embeddings increasingly encode linguistic structure with depth, blending acoustic and linguistic cues \cite{anderson2024}.}


Comparing EF to LF reveals complementary strengths. EF typically improves F1, reflecting superior class separation in a single joint space that captures cross-modal interactions. LF, by averaging calibrated posteriors from independent models, more often achieves lower log loss, indicating better probability calibration. Clinically, maximizing F1 suits screening, whereas minimizing log loss better serves risk estimation and threshold setting, because it requires reliable probabilities.

Related work converges with our observations: Krstev et al. (2022) report EF outperforming LF on dementia-speech classification task \cite{krstev2022}; both Javanmardi et al. (2024) and Pan et al. (2025) find that relying on the last layer underperforms and previous layers may be more informative \cite{javanmardi2024, pan2025}; similarly, Li et al. (2024) present learned layer weighting for Whisper showing results of comparable importance of selected layers to our study \cite{li2024}. \textcolor{black}{Additional layer-wise studies further support the general claim that representation depth controls which factors dominate (signal vs. content) and that the optimal layer is task-dependent \cite{pasad2021, pasad2023, maji2024, cooper2025}, including in other speech and neural modeling settings that explicitly evaluate layer variation effects \cite{anderson2024}.}

\begin{figure}[t!]
 \begin{center}
      \includegraphics[width=0.9\linewidth]{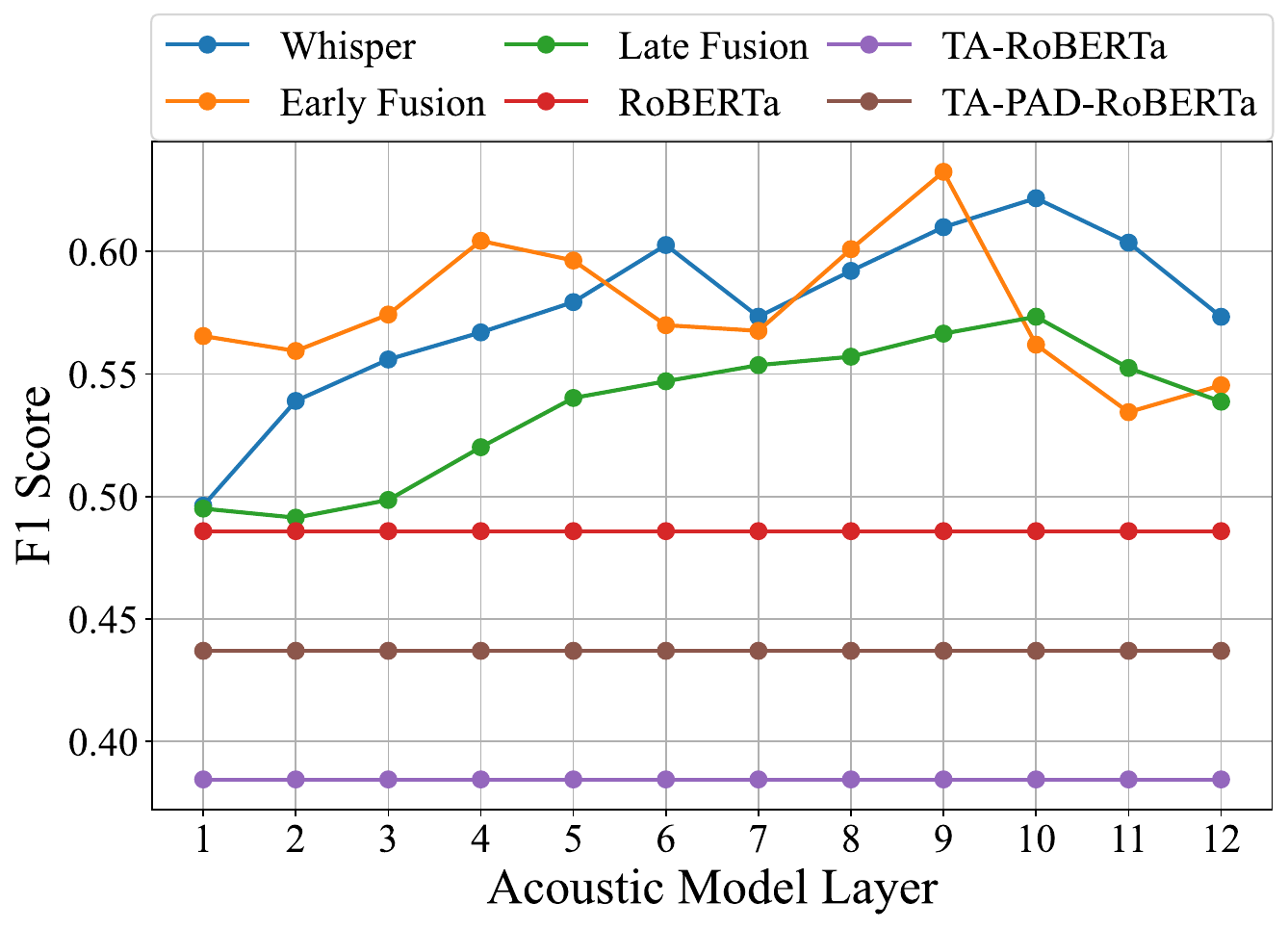} 
    \vspace{-0.1em}
   \vspace{-0.1em}
   \caption{F1 score results of selected strategies depending on the acoustic model layer for Whisper + RoBERTa scenario.}
    \label{fig:f1}
 \end{center}
\vspace{-1em}
\vspace{-1em}
\end{figure}

\section{Conclusion}
\label{sec:conc}
We compared acoustic and linguistic representations together with EF and LF for speech-based cognitive status classification. Mid to late acoustic layers (typically 8–10) consistently performed best across fusion types and in acoustic-only models, supporting their capture of cognitively relevant paralinguistic cues. EF improved discrimination (F1) by modeling joint representations prior to the last acoustic layers, thanks to the complementary coupling of heterogeneous information.

EF maximized accuracy, whereas LF yielded better calibration (lower log loss), implying different clinical roles (classification vs. uncertainty estimation). Text-only variants (TA-, TA-PAD-) underperformed, indicating acoustic signals dominate and simple temporal alignment is insufficient. These findings underscore the importance of fusion design and layer choice; future work should pursue adaptive fusion and richer linguistic grounding to enhance interpretability and clinical utility.

\textcolor{black}{Our aim was not to introduce a new architecture, but to provide a unified, reproducible comparison of acoustic-only, text-only, early/late fusion settings across encoder depth under consistent splits, tuning, and evaluation. The resulting guidance (what modality and what layer to use) targets practical model selection for cognitive screening pipelines.}




\newpage
\section{Acknowledgements}
\label{sec:ackn}

This work was supported by project LangInLife (A Lifetime with Language: The Nature and Ontogeny of Linguistic Communication) no. CZ.02.01.01/00/23\_025/0008726, co-funded by the European Union, and COST Action CA24128 ``European Network to Advance the Development and Implementation of Vocal Biomarkers (eVoiceNet)''.

This study was made possible by a grant from the Fulbright Visiting Student Researcher Program.

}
{
\ninept
\fontsize{9pt}{11pt}\selectfont
\setlength{\bibitemsep}{0.65\baselineskip}

\bibliographystyle{IEEEbib}
\bibliography{strings,refs}
}

\end{document}